\documentclass[11pt]{article}

\usepackage{graphicx}
\usepackage{latexsym}
\usepackage{amstext}
\usepackage{amsxtra}
\newcommand{\bs}{\boldsymbol}

\begin{document}
\title{Vortex lattices in a stirred Bose-Einstein condensate}
\author{K. W. Madison, F. Chevy, W. Wohlleben$^{\dagger}$, J. Dalibard\\
Laboratoire Kastler Brossel$^*$,\\
D\'epartement de Physique de
l'Ecole Normale Sup\'erieure\\
24 rue Lhomond, 75005 Paris, France}
\date{}
\maketitle 

\begin{abstract}
We stir with a focused laser beam a Bose-Einstein condensate
of $^{87}$Rb atoms confined in a magnetic trap.
We observe the formation of a single vortex for a
stirring frequency exceeding a critical value.
At larger rotation frequencies we produce states of the
condensate for which up to eleven vortices are simultaneously present.
We present measurements of the decay of a vortex array
once the stirring laser beam is removed.
\end{abstract}

\noindent {\bf Pacs:} 03.75.Fi, 67.40.Db, 32.80.Lg

\vskip 5mm

\section{Introduction}

The discovery of Bose-Einstein condensation of atomic gases
\cite{Anderson95,Bradley957,Davis95,Fried98} has 
led to a new impulse in the physics of quantum gases.
Among the several questions that can be studied in these
systems, superfluidity is one of the most intriguing and fascinating.
A first hint to the superfluid behaviour of these systems
was provided by the study of the oscillation frequencies
of the normal modes of these systems \cite{Dalfovo99}.
A more direct evidence has been found by recent experiments
aiming to study the energy deposited in the 
condensate by a moving ``object" ({\it i.e.}
the hole created by a blue detuned laser)\cite{Raman99} and
the rotational properties of the gas 
\cite{Matthews99,Madison00,Marago00}. 

\begin{figure}[bht]
\begin{center}
\includegraphics[height=5.58cm,width=3cm]{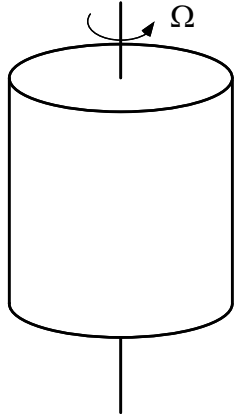}
\end{center}
\caption{\footnotesize The rotating bucket experiment. A fluid
is placed in a cylindrical container, which is rotated at a constant
angular frequency $\bs \Omega$ along its axis. The velocity field
at equilibrium reflects the superfluid character of the gas or the liquid.
Indeed, for a low enough rotation frequency, the superfluid 
is not entrained at all by the
walls of the rotating bucket, and it stays at rest
in the laboratory frame.}
\label{rotbucket}
\end{figure}

The relation between the rotational and superfluid 
properties of a gas or a liquid is illustrated by the famous
 ``rotating bucket'' experiment (see Fig. \ref{rotbucket}).
When an ordinary fluid is placed in a rotating container, 
the steady state corresponds to a rotation of the fluid as a
whole together with the vessel. The velocity field for a
normal fluid is the same for a rigid rotating body:
\begin{equation}
{\bs v}= {\bs \Omega}\times {\bs r}\ .
\label{rigid}
\end{equation}
Superfluidity, first observed in liquid HeII, 
changes dramatically this behaviour 
\cite{Onsager,Feynman,Nozieres,Lifshitz,Donnelly}.
Indeed, for a low enough rotation frequency, the superfluid 
is not entrained at all by the
walls of the rotating bucket, and it stays at rest
in the laboratory frame.

For a more quantitative description of the superfluid phenomenon, consider
a Bose-Einstein condensate with weak, repulsive interactions (see {\it e.g.}
\cite{Nozieres,Lifshitz,Huang}). This model is relevant for the
$^{87}$Rb gas that we investigate experimentally since
(i) the scattering length $a$ characterizing the 
two-body interactions is positive,
(ii) the ``gaseous" parameter $(\bar \rho a^3)^{1/2}$ is much smaller than 1
($\bar \rho$ is the average spatial density of the condensate).
When cooled well below the condensation temperature, nearly
all the atoms occupy the same state, and we can write 
the condensate wave function as 
$\psi(\bs r)= \sqrt{\rho(\bs r)}\; e^{iS(\bs r)}$,
where $\rho(\bs r)$ is the spatial density. The velocity
field is given by $m {\bs v}=\hbar \bs \nabla S$, 
which is manifestly irrotational.
This fact forbids, in particular, the rigid-body velocity field given
in (\ref{rigid}).
An energy analysis, which is outlined in the next section of this paper
for the particular case of harmonic confinement,
leads to the following conclusions.
For a low enough angular rotation frequency $\Omega$,
the ground state is the same as that when the bucket is at rest. 
In this state, the phase of the condensate is
constant over the whole volume and no rotational motion occurs. 
However above a critical frequency $\Omega_c$,
a line of singularity, {\it i.e.} a vortex line, 
appears. The condensate density drops to zero
on axis, and the phase of the condensate wave function
is now given by $S(\bs r)=\theta$, where $\theta$ is the azimuthal
angle around the vortex axis (oriented along $z$). 
The corresponding velocity field varies as $r^{-1}$,
where $r$ is the distance to the vortex center. 
Moreover, the velocity field has a quantized circulation around the axis:
\begin{equation}
\oint {\bs v}\cdot {d \bs r}= \frac{2\pi \hbar}{m}\ ,
\label{quantif}
\end{equation}
where $m$ is the mass of the particles in the fluid.
For frequencies notably higher than $\Omega_c$, one 
might expect that a single vortex line with a quantum number $n$
larger than 1 would appear ($S(\bs r)=n\theta$). However this state is
unstable \cite{Nozieres,Lifshitz,Donnelly} and fragments
into $n$ vortices each with a unit circulation quantum.

These phenomenon have been observed in experiments with
superfluid liquid helium; however, the model of a dilute Bose gas presented
in the previous paragraph is not applicable to this dense system. 
Let us quote here two milestones in this very rich field of research.
In 1958 Vinen performed an experiment
in which he detected the presence of a single vortex filament
by observing the induced frequency shift of the vibrational modes 
of a wire at the center of a rotating liquid helium bath \cite{Vinen58}. 
In 1982 Yarmchuk and Packard obtained images of a vortex lattices 
in the superfluid by imaging electrons initially trapped at the cores of the
vortex lines \cite{Yarmchuk82}.

\section{Vortices in trapped gaseous condensates}

The generation of quantized vortices in trapped atomic gases has been
the subject of numerous theoretical studies in the recent years.  Two schemes
have been considered.  The first one uses laser beams
to engineer the phase of the condensate wave function and produce
the desired velocity field \cite{Marzlin,Dum,Petrosyan,Williams,Dobrek99}.
Recently this scheme \cite{Williams} has been successfully applied to a binary mixture
of condensates, resulting in a quantized rotation of one of the
two components around the second one \cite{Matthews99}.
Phase imprinting has also been used for the generation of solitons inside
a condensate \cite{Burger99,Phillips}.

\begin{figure}[t]
\begin{center}
\includegraphics[height=2.42cm,width=8.7cm]{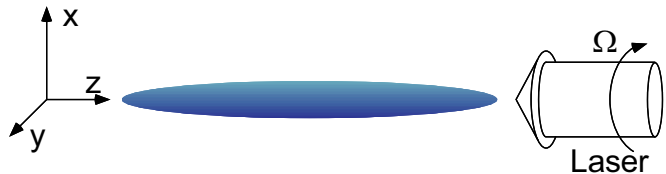}
\end{center}
\caption{\footnotesize A cigar-shaped atomic cloud confined in an axisymetric
magnetic trap is stirred by a far-detuned laser beam. The laser beam propagates
along the long axis of the cigar ($z$), and it creates a dipole potential 
which is anisotropic in the $xy$ plane. This anisotropy rotates around the $z$
axis at the angular frequency $\Omega$.}
\label{stir}
\end{figure}

The second scheme, which is explored in the present work, is 
directly analogous to the rotating bucket experiment \cite{Leggett,Stringari99}.
The atoms are confined in a static, cylindrically-symmetric
Ioffe-Pritchard magnetic trap upon which we 
superimpose a non-axisymmetric,
attractive dipole potential created by a stirring laser beam
(see figure \ref{stir}).
The combined potential leads to a cigar-shaped harmonic trap
with a slightly anisotropic transverse profile.  The
transverse anisotropy is rotated at angular frequency $\Omega$ 
as the gas is evaporatively cooled
to Bose-Einstein condensation, and it plays the role of the bucket wall roughness.

In this scheme, the formation of vortices
is a consequence of thermal equilibrium. 
In the frame rotating at the same
frequency as the anisotropy, the Hamiltonian is time-independent
and one can use a standard thermodynamics approach to 
determine the steady-state of the system.
In this frame, the Hamiltonian can be written 
$\tilde H=H-\Omega L_z$, where $H$
is the Hamiltonian in the absence of rotation,
and $L_z$ is the total orbital angular momentum
along the rotation axis. 
For a gas with repulsive interactions,
the term $-\Omega L_z$ may favor the creation of a
state where the condensate wave function
has an angular momentum $\hbar$ along the $z$ axis
and therefore contains a vortex filament 
\cite{Baym96,Stringari96,Sinha97,Lundh97,Butts99,Feder99,Fetter98,Castin99,Caradoc,Pu}.

\begin{figure}[t]
\begin{center}
\includegraphics[height=3.91cm,width=12.02cm]{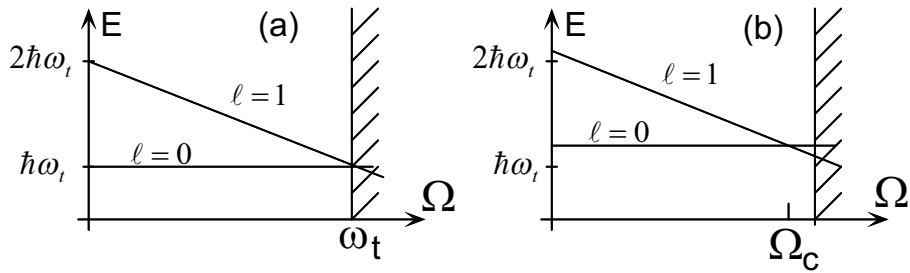}
\end{center}
\caption{\footnotesize
Lowest transverse energy levels for the Gross-Pitaevski equation of a 
 gas confined in a bi-dimensional harmonic potential with frequency $\omega_t$,
and stirred at a angular frequency $\Omega$.
(a) In absence of interaction the ground state of the system is the same
as that for $\Omega=0$ for any stirring frequency in the stability domain 
$\Omega < \omega_t$. (b) For an interacting gas with repulsive 
interactions (assumed here to be weak),
there exists a critical frequency $\Omega_c$ above which the ground state
of the system has an angular momentum $\ell=1$. The effect of the stirring anisotropy
on the energy levels has been neglected.
}
\label{energy}
\end{figure}

This is illustrated in figure \ref{energy}, which displays
the lowest eigenenergies for the Gross-Pitaevski equation
describing  a  gas with $N$ atoms 
confined in a two-dimensional isotropic harmonic potential of frequency $\omega_t$:
\begin{equation}
-\frac{\hbar^2 }{2m}\Delta \psi + \frac{1}{2}m\omega_t^2 (x^2+y^2) \psi
+ \frac{4\pi \hbar^2 N a}{m}|\psi|^2 \psi=E\psi \ .
\label{GP}
\end{equation}
Here $a$ is the scattering length characterizing the 2-body interaction.
In figure \ref{energy}a, we consider an ideal gas ($a=0$). The ground state
found for $\Omega=0$, which has zero angular momentum, remains the 
ground state of the system until the stirring frequency $\Omega$
reaches $\omega_t$. Above this value, the gas is no longer 
stable since the centrifugal force $m\Omega^2 {\bs r}$ exceeds the restoring force of the confining potential
$-m\omega_t^2 \bs r$.
For a gas with repulsive interactions (figure \ref{energy}b), 
the energy of the state with angular
momentum $\ell=1$ crosses the energy of the $\ell=0$ state 
at a critical value $\Omega_c<\omega_\bot$. Indeed one can 
infer that for $\Omega=\omega_\bot$ the energy of the state
$\ell=1$ is strictly below the energy of the state $\ell=0$: (i) 
these two states
have the same energy in absence of interactions for $\Omega=\omega_t$; 
(ii) repulsive interactions increase the energy of these states, 
by an amount which is larger for the $\ell=0$ state, since it 
has a smaller volume than the state with $\ell=1$. 

When the interactions are large enough to reach the Thomas-Fermi limit,
for which the kinetic energy is negligibly 
small compared to the potential and interaction energies, the solution of (\ref{GP}) for $\ell=1$
leads to a radius for the vortex core of the order of 
the healing length $\xi=(8\pi a \rho)^{-1/2}$,
where $\rho$ is the density of the condensate at the center of the trap
in the absence of a vortex \cite{Dalfovo99}.

\section{Experimental setup and determination of $\Omega_c$}

Our set-up has been described in detail previously
\cite{Soding99,Madison00}, and we only briefly outline 
the main elements. The atoms are confined in 
an Ioffe-Pritchard 
magnetic trap. The slow oscillation frequency of the 
elongated magnetic trap is $\omega_z/(2\pi)= 11.7$~Hz ($z$ is horizontal in 
our setup), while the transverse oscillation frequency is $\omega_\bot/(2\pi)=
219$~Hz. Bose-Einstein condensation is reached using 
evaporative cooling produced by a chirped radio-frequency source. 
For the data presented here, the final frequency
of the evaporation ramp is chosen only a few kHz
above the value which completely empties the trap. This garantees
that the temperature of the atomic gas is below 80~nK, the uncondensed
cloud being only marginally visible. The 
number of atoms in the condensate is of the order of 140 000. 

The stirring laser beam is switched on approximately at the time when the
cloud reaches the condensation point.
It propagates along the slow axis of the magnetic trap.
The beam waist is $w_s=20.0\; (\pm\,1)$~$\mu$m and 
the laser power $P$ is 0.4~mW.
The recoil heating induced by this far-detuned beam (wavelength $852$~nm)
is negligible.  Two crossed acousto-optic modulators, combined with 
a proper imaging system, allow for an arbitrary
translation of the laser beam axis with respect to the symmetry axis 
of the condensate.

The motion of the stirring beam
consists in the superposition of a fast and a slow component.
The optical spoon's axis is toggled 
at a high frequency (100~kHz) between two symmetric
positions about the long trap axis $z$.
The intersections of the stirring beam axis and the $z=0$ plane are 
$\pm a (\cos\theta\, {\bf u}_x+\sin \theta \,{\bf u}_y)$,
where the distance $a$ is $8\;\mu$m.
The fast toggle frequency is chosen to be much larger than
the magnetic trap frequencies so that the atoms experience
an effective two-beam, time-averaged potential. 
The slow component of the motion is a uniform rotation of the angle 
$\theta=\Omega t$. The value of the angular frequency $\Omega$ is 
maintained fixed during the evaporation at a value chosen between 0 
and $2\pi\times 220$~rad~${\rm s}^{-1}$.

The dipole potential, proportional to the power of the stirring beam, 
is well approximated by $m\omega_\bot^2 (\epsilon_X X^2 +\epsilon_Y Y^2)/2$.
The ($X,Y$) basis is rotated with respect to the fixed axes ($x,y$) by the
angle $\theta(t)$, and $\epsilon_X=0.03$ and $\epsilon_Y=0.09$ for the 
parameters given above \cite{verif}.  
The action of this beam is essentially a slight modification of 
the transverse frequencies of the magnetic trap
while the longitudinal frequency is nearly unchanged.  
The overall stability of the stirring beam 
on the condensate appears to be a crucial element for the 
success of the experiment, and we estimate that 
our stirring beam axis is fixed to and stable on the condensate 
axis to within 2~$\mu$m. 

After the end of the evaporation ramp, we let the system
reach thermal equilibrium in this ``rotating bucket'' for a duration 
$t_{\rm r}=500$~ms.
The vortices induced in the condensate by the optical spoon
are then studied using a time-of-flight analysis (27~ms) after the atoms
have been released from the magnetic trap.  
Due to the atomic mean field energy, the initial cigar shape
of the atomic cloud transforms into
a pancake shape during the free fall. 
The transverse $xy$ and longitudinal $z$ sizes
grow by a factor of $40$ and $1.2$ respectively \cite{Castin96}. 
In addition, the core size of the vortex should expand at 
least as fast as the transverse size of the condensate 
\cite{Castin96,Lundh98,Dalfovo00}.
Therefore a vortex with an initial diameter
$2\xi=0.4\;\mu$m for our experimental parameters is expected to
grow to a size of 16~$\mu$m.

At the end of the time-of-flight period, we illuminate the atomic 
sample with a resonant probe laser for 20~$\mu$s. 
The shadow of the atomic cloud in the probe beam is imaged 
onto a CCD camera with an optical resolution $\sim 7\,\mu$m.
The probe laser propagates along the $z$-axis 
so that the image reveals the column density of the cloud after expansion
along the stirring axis.

\begin{figure}[t]
\begin{center}
\includegraphics[height=4.03cm,width=10.53cm]{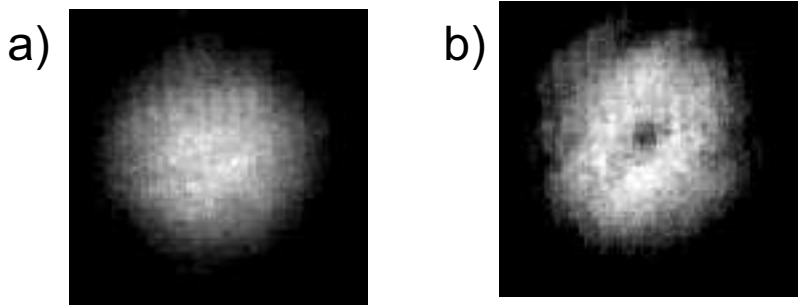}
\end{center}
\caption{\footnotesize
Absorption images of a Bose-Einstein condensate
stirred with a laser beam (after a 27~ms time-of-flight).
For both images, the condensate number is $N_0=(1.4\;\pm 0.5)\;10^{5}$
and the temperature is below 80~nK.
The rotation frequency $\Omega/(2\pi)$ is respectively
(a) 145~Hz; (b) 152~Hz.}
\label{vort01}
\end{figure}

For the trap parameters given above, we have found that no singularity
appears in the cloud image as long as the stirring frequency is
smaller than $149$~Hz. However, for a stirring frequency
equal to or above 150~Hz (but smaller than 160~Hz, see next section)
a density dip systematically appears at the center of the cloud.
Figure \ref{vort01} illustrates this transition by showing 
two pictures of the condensate after expansion;
the first with a stirring frequency 
$\Omega=145$~Hz (Fig.\ref{vort01}a) below and the second
$\Omega=152$~Hz (Fig.\ref{vort01}b) above the critical frequency.

The dip in the optical density can reach up to 50\,\% of the maximal 
column density
and it constitutes an unambiguous signature of the
presence of a vortex filament. In particular it cannot be
induced by a mechanical action of the stirring laser, 
which in fact creates a restoring force towards the center
of the magnetic trap since it is detuned red of resonance.
The fact that the density does not vanish in the dip may
be due to several reasons: (i) The vortex filament 
may not be perfectly straight but rather an oscillating line
(the so-called Thomson mode \cite{Lifshitz}). 
(ii) Some non-condensed atoms may be trapped
in the core of the vortex. (iii) The core diameter
after expansion of $\sim 18\,\mu$m at the half max of the dip is
not much larger than the resolution limit of the imaging optics.

Finally, we note that the critical frequency is notably 
larger than the predicted value
of 91~Hz \cite{Lundh97} (see also 
\cite{Baym96,Sinha97,Castin99,Feder99}).
This deviation may be due to the marginality 
of the Thomas-Fermi approximation
for our relatively low condensate number.

\section{Dynamics of a vortex array}

When the stirring frequency is increased notably above the value
$\Omega_c$, multiple dips appear in the absorption image of the cloud. 
Each of these dips has approximately the same spatial width as
the single dip of Fig. \ref{vort01}.  For the
trap parameters given above, the threshold frequency for the
appearance for such multi-vortex patterns is 160~Hz, and 
condensates with up to 4 vortices have been observed \cite{Madison00}.

We have found that we can generate patterns with more than 
4 vortices if we reduce the transverse trapping frequency.
This can easily be done by adjusting the
value of the bias magnetic field at the center of the trap. The data
that we give now have been obtained with a transverse
frequency of $\omega_t\;/\;2\pi= 169$~Hz and a longitudinal frequency 
of $\omega_z\;/\;2\pi= 11.7$~Hz, the same as before.

\begin{figure}[tbh]
\begin{center}
\includegraphics[height=4.76cm,width=14.19cm]{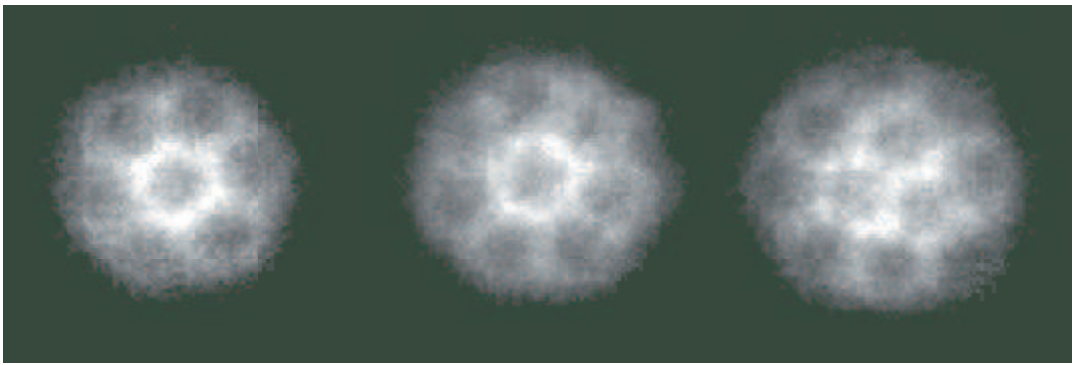}
\end{center}
\caption{\footnotesize Arrays of vortices in a Bose-Einstein
condensate stirred by a laser beam (after a 27~ms time-of-flight).
These pictures have been obtained with a magnetic trap less stiff
than for the pictures of Fig.\ref{vort01} ($\omega_t\;\ \; 2\pi=169$~Hz). 
}
\label{vort78}
\end{figure}

For an atom number similar to the previous data ($\sim\;150\,000$),
the critical frequency is now $\Omega_c\;/\;2\pi= 119$~Hz. This
still corresponds to a ratio $\Omega_c/\omega_t \sim 0.7$. When we increase
the stirring frequency above 130~Hz,  arrays of several vortices 
can be observed.
For instance, Fig.\ref{vort78} shows images of condensates with 7, 8 and 11
vortices. These three images have all been obtained with the same experimental
conditions ($\Omega\;/\;2\pi= 135$~Hz), 
apart from residual shot-to-shot fluctuations.
In the pictures with 7 and 8 vortices (Fig.\ref{vort78}a and Fig.\ref{vort78}b), 
one vortex is found at the center
of the trap, and the remaining ones form a regular hexagon or heptagon.
The 11-vortex pattern (Fig.\ref{vort78}c) has 3 vortices forming
an equilateral triangle surrounded by a regular octogon. Some of the dips
of the octogon are hardly visible, since they are located at the border
of the condensate cloud.

In \cite{Madison00} we studied the lifetime of the one vortex pattern
of Fig.\ref{vort01}b when the stirring laser is removed. We found that this
structure may survive for up to $\sim$~1~s before decaying, proving 
the metastability of the current associated with the vortex \cite{Rokhsar97}. We now
present similar data obtained for a multiple vortex structure.
For this experiment we chose a rotation frequency of $\Omega=135$~Hz
and adjusted the number of atoms in the condensation to maximize the
probability that a 5-vortex pattern form.
For each point in the decay study, a vortex lattice is formed,
then the stirring beam is turned off, and the condensate is allowed
to evolve in the pure magnetic trap for an adjustable time.
A time-of-flight analysis is performed to determine 
the number of vortices still present.

\begin{figure}[tbh]
\begin{center}
\includegraphics[height=5.29cm,width=12.78cm]{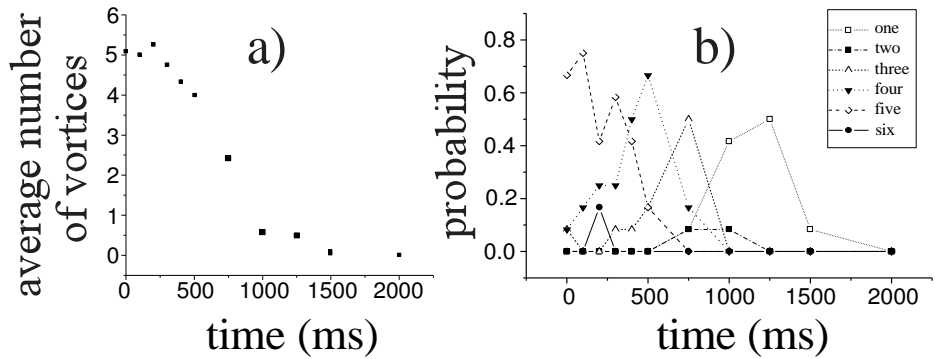}
\end{center}
\caption{\footnotesize Decay of a vortex array. (a) Average number of vortices
as a function of time spent by the gas in the axisymetric trap after the
end of the stirring phase. (b) Fraction of the images displaying $n$
vortices as a function of time. After 2 seconds all pictures exhibit a condensate with no 
vortex.
}
\label{lifetime}
\end{figure}

The results are displayed in Fig.\ref{lifetime}
which presents the decay of the average 
number of vortices as a function of time, each point representing 
the average of 10 shots. The characteristic time for the 
reduction by a factor 2 of this average number is 750~ms.
A more detailed representation is shown in
Fig.\ref{lifetime}b which displays the fraction of pictures showing
$n$ vortices ($n=0,\ldots,6$) as a function of time.
It seems apparent that the vortex pattern decays
by losing one vortex at a time. 
For instance, at time $t=100$~ms, 9 patterns out of 10
exhibit 5 vortices; 400 ms later, 7 patterns out of 10 exhibit 4 vortices. 
It is remarkable that the vortex pattern
readjusts itself shortly after the loss of a vortex. 
For instance, the 3-vortex patterns are
found most often as a quasi-centered and quasi-equilateral triangle, 
while the 4-vortex patterns have essentially a square shape.

\section{Conclusion}

We have reported the formation of single and multiple-vortex 
structures in a gaseous Bose-Einstein condensate when it is 
stirred by a laser beam which produces a
slight rotating anisotropy. 
We have also presented measurements of the lifetime of the
pattern of vortices when the rotating anisotropy is removed.

Several extensions of this work can be considered.
Direct evidence for the quantized circulation of the velocity
field (Eq.\ref{quantif}) could be obtained either from 
an interferometric measurement similar to the one performed
in \cite{Raman99} (see {\it e.g.} \cite{Castin99} or \cite{Dobrek99}), 
or from a study of the elementary excitations of the vortex
filament \cite{Dodd97,Zambelli98,Svidzinsky}.
Also the role of the thermal component in the nucleation and
decay of the vortex pattern remains to be elucidated
\cite{Feder99,Castin99,Machida99,Svidzinsky98A,Fedichev99}.
The rotational properties of the gas before the condensation
point, where the collisional dynamics can be described 
by the classical or quantum Boltzmann equation, constitute
other very interesting questions \cite{dgo00}. Finally,
the scattering of quasi-resonant light by this medium
may lead to spectacular phenomena, such as a 
``black hole" type behaviour when the speed of sound
or mass flow in the condensate exceeds that of light \cite{BH}.

We wish to dedicate this paper to the memory of Dan Walls who has 
had a seminal influence on the field of quantum optics in general, and in 
the understanding of the physics of gaseous Bose Einstein condensates in 
particular. 

\noindent {\bf Acknowledgments:}
we thank V. Bretin, Y. Castin, C. Cohen-Tannoudji, C. Deroulers,
D. Gu\'ery-Odelin, C. Salomon,
G. Shlyapnikov, S. Stringari, and the ENS Laser cooling
group for several helpful discussions and comments. 
This work was partially supported by CNRS, Coll\`{e}ge de France,
DRET, DRED and EC (TMR network ERB FMRX-CT96-0002). This material is
based upon work supported by the North Atlantic Treaty Organization
under an NSF-NATO grant awarded to K.M. in 1999.

$^\dagger$ permanent address: Max Planck Institute f\"ur KernPhysik, Heidelberg, Germany.

$^*$ Unit\'e de Recherche de l'Ecole normale sup\'erieure et de
l'Universit\'e Pierre et Marie Curie, associ\'ee au CNRS.


\begin{thebibliography}{}


\bibitem{Anderson95} M. H. Anderson, 
J. Ensher, M. Matthews, C. Wieman, and E. Cornell, 
Science {\bf 269}, 198 (1995).

\bibitem{Bradley957} C. C. Bradley, C. A. Sackett, and R. G. Hulet, Phys. Rev. Lett. {\bf 78}, 985 (1997); see also C. C. Bradley, C. A. Sackett, J. J. Tollett, and R. G. Hulet, Phys. Rev. Lett. {\bf 75}, 1687 (1995).

\bibitem{Davis95} K. B. Davis, M.O. Mewes, N. Van Druten, D. Durfee,
D. Kurn, and W. Ketterle, 
Phys. Rev. Lett. {\bf 75}, 3969 (1995).

\bibitem{Fried98} D. Fried, 
T. Killian, L. Willmann, D. Landhuis, S. Moss, 
D. Kleppner, and T. Greytak, 
Phys. Rev. Lett. {\bf 81}, 3811 (1998).

\bibitem{Dalfovo99} For a review, see {\it e.g.} 
A. S. Parkins and D. Walls, Phys. Rep. {\bf 303}, 1 (1998); 
F. Dalfovo, S. Giorgini, L. P. Pitaevskii, and S. Stringari, Rev. Mod. Phys. {\bf 71}, 463 (1999).

\bibitem{Raman99} C. Raman, M. Köhl, R. Onofrio, D. S. Durfee, C. E. Kuklewicz, Z. Hadzibabic, and W. Ketterle, Phys. Rev. Lett. {\bf 83}, 2502 (1999).

\bibitem{Matthews99} M. R. Matthews, B. P. Anderson, P. C. Haljan, D. S. Hall, C. E. Wieman, and E. A. Cornell, Phys. Rev. Lett. {\bf 83}, 2498 (1999).

\bibitem{Madison00} K. Madison, F. Chevy, W. Wohlleben, and J. Dalibard,
Phys. Rev. Lett. {\bf 84}, 806  (2000).

\bibitem{Marago00} O. M. Marag\`o, 
S. A. Hopkins, J. Arlt, E. Hodby, G. Hechenblaikner, and C. J. Foot,
Phys. Rev. Lett. {\bf 84}, 2056 (2000).

\bibitem{Onsager} L. Onsager, Nuovo Cimento {\bf 6}, suppl. 2, 249 (1949).

\bibitem{Feynman} R.P. Feynman, in ``Progress in Low Temperature Physics", vol. 1,
ed. by C.J. Gorter (North-Holland, Amsterdam, 1955) Chapter 2.

\bibitem{Nozieres} P. Nozi\`eres and D. Pines, {\it The Theory of Quantum Liquids},
vol. II (Addison Wesley, Redwood City, 1990).

\bibitem{Lifshitz} 
E.M. Lifshitz and L. P. Pitaevskii, {\it 
Statistical Physics, Part 2}, chap. III (Butterworth-Heinemann, 
Oxford, 1980).

\bibitem{Donnelly}
R.J. Donnelly, {\it Quantized Vortices in Helium II}, (Cambridge, 1991).

\bibitem{Huang} K. Huang, {\it Statistical Mechanics}, John Wiley (New-York, 1963).

\bibitem{Vinen58}
W.F. Vinen, Nature {\bf 181}, 1524 (1958) and Proc. Roy. Soc. A {\bf 260}, 218 (1961).

\bibitem{Yarmchuk82} E. J. Yarmchuck and R. E. Packard, J. Low Temp. Phys. {\bf 46}, 479 (1982).

\bibitem{Marzlin} K.P. Marzlin and W. Zhang, Phys. Rev. Lett. {\bf 79}, 4728 (1997).

\bibitem{Dum} R. Dum, J. I. Cirac, M. Lewenstein, and P. Zoller, Phys. Rev. Lett. {\bf 80}, 2972 (1998).

\bibitem{Petrosyan} K.G. Petrosyan and L. You, Phys. Rev. A {\bf 59}, 639 (1999).

\bibitem{Williams} J. Williams and M. Holland, Nature {\bf 401}, 568 (1999).

\bibitem{Dobrek99} L. Dobrek, M. Gajda, M. Lewenstein, K. Sengstock, G. Birkl, and W. Ertmer, Phys. Rev. A {\bf 60}, R3381 (1999).

\bibitem{Burger99} S. Burger, K. Bongs, S. Dettmer, W. Ertmer, K. Sengstock,
A. Sanpera, G. V. Shlyapnikov, and M. Lewenstein, 
Phys. Rev. Lett. {\bf 83}, 5198 (1999).

\bibitem{Phillips} J. Denschlag, J.E. Simsarian, D.L. Feder, 
C. Clark, L.A. Collins, 
J. Cubizolles, L. Deng, E.W. Hagley, K. Helmerson, W.P. Reinhardt, S.L. Rolston, 
B.I. Schneider, and W.D. Phillips,  Science {\bf 287}, 97 (2000). 

\bibitem{Leggett} A.J. Leggett, Topics in Superfluidity and superconductivity, in 
{\it Low Temperature Physics}, Edts M. Hoch and R. Lemmer (Springer-Verlag, 1992).

\bibitem{Stringari99} S. Stringari, Phys. Rev. Lett. {\bf 82}, 4371 (1999).

\bibitem{Baym96} G. Baym and C.J. Pethick, Phys. Rev. Lett. {\bf 76}, 6 (1996).

\bibitem{Stringari96} F. Dalfavo and S. Stringari, Phys. Rev. A {\bf 53}, 2477 (1996).

\bibitem{Sinha97} S. Sinha, Phys. Rev. A {\bf 55}, 4325 (1997).

\bibitem{Lundh97} E. Lundh, C. J. Pethick, and H. Smith, Phys. Rev. A {\bf 55}, 2126 (1997).

\bibitem{Butts99} D. Butts and D. Rokshar, Nature {\bf 397}, 327 (1999). 

\bibitem{Feder99} D. L. Feder, C. W. Clark, and B. I. Schneider, Phys. Rev. Lett. {\bf 82}, 4956 (1999).

\bibitem{Fetter98} A. Fetter, J. Low. Temp. Phys. {\bf 113}, 189 (1998). 

\bibitem{Castin99} Y. Castin and R. Dum, Eur. Phys. J. D. {\bf 7}, 399 (1999).

\bibitem{Caradoc} B. M. Caradoc-Davies, R. J. Ballagh and K. Burnett, Phys. Rev. Lett. {\bf 83}, 895 (1999).

\bibitem{Pu} H. Pu, C. K. Law, J. H. Eberly and N. P. Bigelow, Phys. Rev. A {\bf 59}, 1533 (1999).


\bibitem{Soding99} J. S\"oding, D. Gu\'ery-Odelin, P. Desbiolles, F. Chevy, H. Inamori, J. Dalibard, Appl. Phys. B {\bf 69}, 257 (1999).

\bibitem{verif} We observed that 
the vortex formation is not sensitive to the exact value 
of $\epsilon_X,\epsilon_Y$, and a single vortex state under the
same conditions was successfully created with a laser power 
of half and twice the value of 0.4~mW used here.

\bibitem{Castin96} Y. Castin and R. Dum, Phys. Rev. Lett. {\bf 77}, 5315 (1996).

\bibitem{Lundh98} E. Lundh, C. J. Pethick, and H. Smith, Phys. Rev. A {\bf 58}, 4816 (1998).

\bibitem{Dalfovo00} F. Dalfovo and M. Modugno, cond-mat/9907102.

\bibitem{Rokhsar97} D. Rokhsar, Phys. Rev. Lett. {\bf 79}, 2164 (1997). 

\bibitem{Dodd97} R. Dodd, K. Burnett, M. Edwards, C. W. Clark, Phys. Rev. A {\bf 56}, 587 (1997).

\bibitem{Zambelli98} F. Zambelli and S. Stringari, Phys. Rev. Lett. {\bf 81},
1754 (1999).

\bibitem{Svidzinsky} A. Svidzinsky and A. Fetter, Phys. Rev. A. {\bf 58}, 3168 (1998).

\bibitem{Machida99} T. Isoshima and K. Machida, Phys. Rev. A {\bf 60}, 3313 (1999).

\bibitem{Svidzinsky98A} A. Svidzinsky and A. Fetter, cond-mat/9811348.

\bibitem{Fedichev99} P. Fedichev and G. Shlyapnikov, Phys. Rev. A {\bf 60},
R1779 (1999).


\bibitem{dgo00} D. Guery-Odelin, cond-mat/0003024

\bibitem{BH} U. Leonhardt and P. Piwnicki, Phys. Rev. Lett. {\bf 84}, 822 (2000).

\end{thebibliography}
\end{document}